\documentclass[letterpaper,twocolumn,10pt]{article}
\usepackage{usenix2019_v3}
\usepackage{algorithm}
\usepackage{algorithmic}
\usepackage{multirow}

\usepackage{tikz}
\usepackage{amsmath}
\usepackage{cleveref}
\crefformat{section}{\S#2#1#3} % see manual of cleveref, section 8.2.1
\crefformat{subsection}{\S#2#1#3}
\crefformat{subsubsection}{\S#2#1#3}
\usepackage{float}
\usepackage{comment}

\usepackage{filecontents}

\begin{document}
%-------------------------------------------------------------------------------

%don't want date printed
\date{}

\title{\Large \bf NetReduce: RDMA-Compatible In-Network Reduction \\
	for Distributed DNN Training Acceleration}

%for single author (just remove % characters)

\author{
{\rm Shuo Liu\footnotemark[1]}\\ 
Huawei
\and
{\rm Qiaoling Wang\footnotemark[1]}\\
Huawei
\and
{\rm Junyi Zhang}\\
Huawei
\and
{\rm Qinliang Lin}\\
Huawei
\and
{\rm Yao Liu}\\
CityU of HK
\and
{\rm Meng Xu}\\
Huawei
\and
{\rm Ray C.C. Cheung}\\
CityU of HK
\and
{\rm Jianfei He}\\
Huawei
} % end author

\maketitle

\renewcommand{\thefootnote}{\fnsymbol{footnote}}
\footnotetext[1]{Equal contribution.}
%-------------------------------------------------------------------------------
\begin{abstract}
%-------------------------------------------------------------------------------
We present NetReduce, a novel RDMA-compatible in-network reduction architecture to accelerate distributed DNN training. Compared to existing designs, NetReduce maintains a reliable connection between end-hosts in the Ethernet and does not terminate the connection in the network. The advantage of doing so is that we can fully reuse the designs of congestion control and reliability in RoCE. In the meanwhile, we do not need to implement a high-cost network protocol processing stack in the switch, as IB does. 

The prototype implemented by using FPGA is an out-of-box solution without modifying commodity devices such as NICs or switches. For the coordination between the end-host and the switch, NetReduce customizes the transport protocol only on the first packet in a data message to comply with RoCE v2. The special status monitoring module is designed to reuse the reliability mechanism of RoCE v2 for dealing with packet loss. A message-level credit-based flow control algorithm is also proposed to fully utilize bandwidth and avoid buffer overflow.

We study the effects of intra bandwidth on the training performance in multi-machines multi-GPUs scenario and give sufficient conditions for hierarchical NetReduce to outperform other algorithms. We also extend the design from rack-level aggregation to more general spine-leaf topology in the data center. NetReduce accelerates the training up to 1.7x and 1.5x for CNN-based CV and transformer-based NLP tasks, respectively. Simulations on large-scale systems indicate the superior scalability of NetReduce to the state-of-the-art ring all-reduce.

\end{abstract}

%-------------------------------------------------------------------------------
\section{Introduction}
%-------------------------------------------------------------------------------

Over the last decade, numerous Artificial Intelligent (AI) applications, such as computer vision (CV)~\cite{krizhevsky2012imagenet,simonyan2014very,he2016deep} and natural language processing (NLP)~\cite{sutskever2014sequence,liu2016recurrent}, have benefited from the rapid development of Deep Neural Networks (DNNs). However, DNN training remains time-consuming due to the growing size of training datasets and models. Since 2012, the computing requirement of AI training has been increasing exponentially with a 3.4-month doubling time~\cite{openai2018ai}, which is much faster than the observation described by Moore's Law (i.e., 2-year doubling period). Therefore, the scale-up techniques~\cite{jouppi2017datacenter,nvidia2020v100,habana2019gaudi} concentrating on the computing capability of a single device cannot fulfill the requirement. In this paper, we focus on the scale-out strategy, distributed system, for DNN training.

Distributed DNN training with synchronous data parallelism is commonly employed in practice. Each computing node has an entire model replica and cannot iterate until the model parameters are synchronized. Unfortunately, this is increasingly a network-bound workload since communication becomes a bottleneck at scale~\cite{daily2018gossipgrad,luo2018parameter,shi2019mg}. The extra communication overhead caused by parameter synchronization makes it difficult for the system to achieve linear scaling. Moreover, the commonly used communication strategies have low efficiency. For example, the approaches based on Parameter Server (PS) ~\cite{dean2012large,li2014scaling,chilimbi2014project} easily lead to the incast of network traffic and wasted resources. All-reduce approaches~\cite{gibiansky2017bringing,sergeev2018horovod} decentralize the workload by using a peer-to-peer communication pattern but require to transmit the data with twice the original model size (\cref{subsec:sync}). 

Recently, a new direction to accelerate distributed DNN training by using in-network aggregation has been explored~\cite{sapio2019scaling,graham2020scalable}. Compared to the state-of-the-art all-reduce strategy, this approach reduces nearly half the aggregation data by offloading gradients aggregation from end-hosts to the network switch (\cref{subsec:int}). Two important issues arise upon practical implementation of in-network aggregation. First, the network should provide efficient and robust transport to fulfill the high requirement of end-to-end throughput in distributed DNN training. Second, in-network aggregation as an add-on function of the switch should be implemented considering the balance between the induced cost and the processing efficiency in the critical forwarding path.   

Unfortunately, existing designs~\cite{sapio2019scaling,graham2020scalable} are far from the above-mentioned requirements. SwitchML~\cite{sapio2019scaling} uses the User Datagram Protocol (UDP) for network transport which cannot provide congestion control of the traffic. When facing packet loss, the system solely relies on application-layer timeout to trigger retransmission which introduces extra latency. Additionally, SwitchML uses programmable switch Application-Specific Integrated Circuit (ASIC), Tofino~\cite{barefoot2019tofino}, to implement in-network aggregation. Such ASIC chip is incapable of processing full-length Ethernet frames due to its limited register resource in the match-action pipeline and thus results in an end-to-end throughput reduction. SHARP~\cite{graham2020scalable}, on the contrary, employs reliable connection (RC) transport by using Remote Direct Memory Access (RDMA) technology, InfiniBand (IB)~\cite{infiniband2015infiniband}, which is more robust. However, instead of those for existing Ethernet, IB relies on specific hardware (e.g., host channel adapter, new optical module) to establish connections between the end-host and the network switch, which is not cost-effective. In this paper, we utilize RDMA over Converged Ethernet (RoCE) protocol which encapsulates the IB transport packet over Ethernet for balancing the robustness and the cost of the in-network aggregation system. Specifically, we focus on RoCE v2~\cite{infiniband2014rocev2} which exists on top of either IPv4 or IPv6 protocol. We will further describe the details in \cref{subsec:roce}. 

In this paper, we present NetReduce, an RDMA-compatible in-network reduction solution to accelerate distributed DNN training. Unlike that existing designs offload the PS into the network directly, NetReduce maintains a reliable connection between end-hosts in the Ethernet and does not terminate the connection in the network. The advantage of doing so is that we can fully reuse the designs of congestion control and reliability in RoCE v2. In the meanwhile, we do not need to implement a high-cost network protocol processing stack in the switch, as IB does. 

We implement NetReduce via Field Programmable Gate Arrays (FPGA), without any modification to commodity Ethernet Network Interface Controller (NIC) or switch. In SwitchML, a single packet carries only 128 bytes of gradients which cannot fully exploit the packet processing capability of the NIC and the switch. On the contrary, NetReduce by using FPGA is capable of processing a packet with a payload size of 1024 bytes, increasing the end-to-end throughput. Very little previous work aims to optimize the aggregation of the long message which faces the challenge of addressing a huge amount of data in a single collective operation. For example, in SHARP~\cite{graham2016scalable}, a piece of the message only contains one packet. Nevertheless, the NetReduce switch can handle the message with 170 KB, which introduces less than 3 $\mu$s extra Round-Trip Time (RTT) latency with additional FPGA operations.

The contributions of this paper are summarized as follows:

1. We present NetReduce, the first implementation of RoCE-compatible in-network aggregation in the field, to the best of authors' knowledge. NetReduce cost-effectively realizes in-network aggregation by not terminating the end-to-end connection. A recovery algorithm (\cref{subsubsec:recovery}) is proposed and a status monitoring module (\cref{subsubsec:pktloss}) is designed to support native RoCE v2 protocol. We also extend the design from rack-level aggregation to more general spine-leaf topology in the data center (\cref{subsec:clos}).

2. At the host side, we design a new transport layer protocol to coordinate end-hosts with the switch (\cref{subsec:protocol}). We also propose a message-level credit-based flow control algorithm to fully utilize bandwidth and avoid the overflow of switch buffer (\cref{subsec:flowcontrol}). 

3. We develop communication cost models of different training algorithms in multi-machines multi-GPUs scenario. Based on the developed models, we give sufficient conditions that hierarchical NetReduce is more communication-efficient than other algorithms (\cref{subsec:mulgpu}).

4. We implement a NetReduce prototype by using FPGA. The experimental results in both multi-machines single-GPU (\cref{subsec:eval_mmsg}) and multi-machines multi-GPUs (\cref{subsec:eval_mmmg}) cases show that NetReduce provides a larger data processing throughput than traditionally used algorithms. Additionally, NetReduce with fixed-point arithmetic does not impact the training convergence compared with floating-point arithmetic. 

5. We perform simulations based on the aforementioned models to evaluate the communication cost of different algorithms, in large-scale distributed DNN training systems involving up to thousands of GPUs. The simulation results are consistent with previous research and indicate the superior scalability of NetReduce to ring all-reduce (\cref{subsec:simulation}).

%-------------------------------------------------------------------------------
\section{Background and Motivation}\label{sec:background}
%-------------------------------------------------------------------------------

\subsection{Parameter Synchronization}\label{subsec:sync}

In general, methods of parameter synchronization in distributed DNN training can be classified into two categories: PS and all-reduce. The PS-based approaches generally work in a ``push + pull'' way. At each iteration, all workers first push their computed gradients to PSs. Then PSs aggregate the gradients and update the model with new weights. Finally, workers pull the updated weights from PSs and start the next computing iteration. PS can easily become a bottleneck: the push phase leads to the traffic incast and the pull phase results in data redundancy in the network. Moreover, the benefits by using PSs depend on the additional CPU resources provided~\cite{awan2017s,bytedance2019byteps}. Assume a homogeneous training system (i.e., all the machines are equipped with the same number of GPUs), the PS does not save anything.

An all-reduce operation performs reductions on data across nodes and writes the result to each node. This process consists of two phases: scatter-reduce and all-gather, where each node ends up with partial and global aggregation data, respectively. An all-reduce operation can be implemented in different ways. A halving/doubling algorithm is employed in~\cite{goyal2017accurate} which has two major drawbacks: 1) the overhead of data transfer is doubled for non-power-of-two case~\cite{thakur2005optimization}; 2) the communication pattern involved may lead to network contention~\cite{patarasuk2009bandwidth}. Baidu~\cite{gibiansky2017bringing} introduces ring all-reduce which has become the most popular algorithm ever since. This implementation is bandwidth-optimal since contention-free communication can be achieved.

\subsection{In-Network Aggregation}\label{subsec:int}

In-network aggregation accelerates distributed DNN training by offloading gradients aggregation into the network switch. For example, SwitchML~\cite{sapio2019scaling} shows that using a programmable data plane~\cite{bosshart2013forwarding,barefoot2019tofino,jin2017netcache,jin2018netchain,sapio2017network} to aggregate gradients on-path reduces the amount of data transferred. SHARP~\cite{graham2016scalable,graham2020scalable} is an IB-compatible hardware architecture for in-network aggregation which relies on fixed-function ASIC. In addition to DNN, iSwitch proposed in~\cite{li2019accelerating} accelerates reinforcement learning which generates more frequent gradient aggregations with smaller sizes by using NetFPGA~\cite{netfpga}.

Suppose that a homogeneous distributed DNN training system has $P$ ($P \geq 2$) GPUs and each machine is equipped with one GPU. To synchronize data with size $M$ by using ring all-reduce, each node splits the data $P$ pieces and transmits a data block with a size of $\frac{M}{P}$ at each step. For example, in Figure~\ref{fig:framework}(A), Node 0 sends a data block with $\frac{M}{P}$ to Node 1 and simultaneously receives a different data block with the same size from Node 3. The procedure completes in $2(P-1)$ steps with $\frac{2(P-1)}{P}M$ amount of data transmitted per node. As $P$ increases, this amount is nearly twice the original model size.

The time taken to complete a ring all-reduce operation can be modeled~\cite{thakur2005optimization} as
\begin{equation}
T_{ring}=2(P-1)\alpha + \frac{2(P-1)}{P}\frac{M}{B}
\label{eq:ring}
\end{equation} 
where $\alpha$ is the latency per message independent of $M$, including the time taken for data preparation and sending interface calls, etc.; $B$ is the network bandwidth. 

\begin{comment}
Eq.(\ref{eq:ring}) consists of two items: the processing latency with $\alpha$ and the tensor transmission time with $M$. The $\alpha$ item relates to the number of transmissions and accounts for the latency by the end-host of processing $2(P-1)$ pieces of the message. The $M$ item accounts for the time that needs to transmit the data with size $M$ under port rate $B$.      
\end{comment}

Similarly, the communication cost of NetReduce can be modeled as
\begin{equation}
T_{inet}=\alpha + \frac{M}{B}
\label{eq:netreduce}
\end{equation}
Compared to ring all-reduce, the communication cost of in-network aggregation is independent of the number of nodes $P$. Unlike that ring all-reduce transmits the message $2(P-1)$ times, in-network aggregation only transmits once, reducing the complexity from $O(P)$ to $O(1)$. Additionally, the data amount transmitted by each node is reduced from $\frac{2(P-1)}{P}M$ to $M$, by nearly 50$\%$ as $P$ increases.

Eq.(\ref{eq:ring}) subtracting Eq.(\ref{eq:netreduce}) gives
\begin{eqnarray}
\Delta T = T_{ring}-T_{inet} = (2P-3)\alpha + \frac{P-2}{P}\frac{M}{B}
\label{eq:subtract_single1}
\end{eqnarray}
When $P \geq 2$, Eq.(\ref{eq:subtract_single1}) $>$ 0, which means that in-network aggregation always takes less communication time than ring all-reduce in the multi-machines single-GPU case.

\subsection{Why RoCE Matters}\label{subsec:roce}

SwitchML~\cite{sapio2019scaling} is built upon UDP for network transport. One issue of using UDP lies in the I/O performance since using UDP solely cannot fully use network bandwidth, e.g., 100 Gbps. Therefore, SwitchML employs Data Plane Development Kit (DPDK)~\cite{dpdk2020} to bypass the kernel for increasing port throughput (i.e., packets per second). To achieve full 100 GE bandwidth, many CPU cores need to be bound to a specific port by using DPDK, which wastes lots of CPU cycles. We aim at the real-world Data Center Network (DCN) environment where CPU resources matter. 

Another issue is that UDP cannot handle transport-level congestion control and retransmission. In typical distributed DNN training jobs, network transportation is in RC mode. However, SwitchML relies on application-level timeout to deal with packet loss which again occupies extra CPU cycles. This may not be a severe problem if a clean slate network dedicated to AI training is assumed. But in a real DCN environment with tremendous background traffic, only application-level flow control is not enough.

The state-of-the-art DCN employs RoCE v2 protocol for transportation~\cite{guo2016rdma}. RoCE v2 increases I/O throughput by a kernel-bypassing technique which reduces CPU overhead and processing latency. Additionally, the NIC supporting RoCE has a complete mechanism of congestion control and reliability assurance.

In terms of low latency, IB suits well in SHARP, which aggregates relatively-small messages/packets in High-Performance Computing (HPC) scenario~\cite{graham2016scalable}. However, distributed DNN training requires high throughput where RoCE v2 does not lose to IB. More importantly, RoCE v2 supports a larger network in a more cost-effective manner than IB. SHARP terminates the end-to-end connection in the network by using special devices such as Target Channel Adapter (TCA), which is not compatible with ordinary Ethernet. On the contrary, NetReduce as an add-on function that supports RoCE v2 does not modify Ethernet commodity devices such as NIC or switch. 

%-------------------------------------------------------------------------------
\section{NetReduce Overview}\label{sec:overview}
%-------------------------------------------------------------------------------

In this section, we first describe the fundamental operation principle of NetReduce and then present a theoretical analysis of communication cost in comparison to other approaches. For easy understanding, we start to introduce NetReduce from a simple multi-machines single-GPU scenario (i.e., each machine is equipped with only one GPU) before discussing the more general case.
 
%-------------------------------------------------------------------------------
\subsection{Single GPU in A Machine}\label{subsec:onegpu}
%-------------------------------------------------------------------------------
 
\begin{figure}[t]
	\centerline{\includegraphics[scale=0.4]{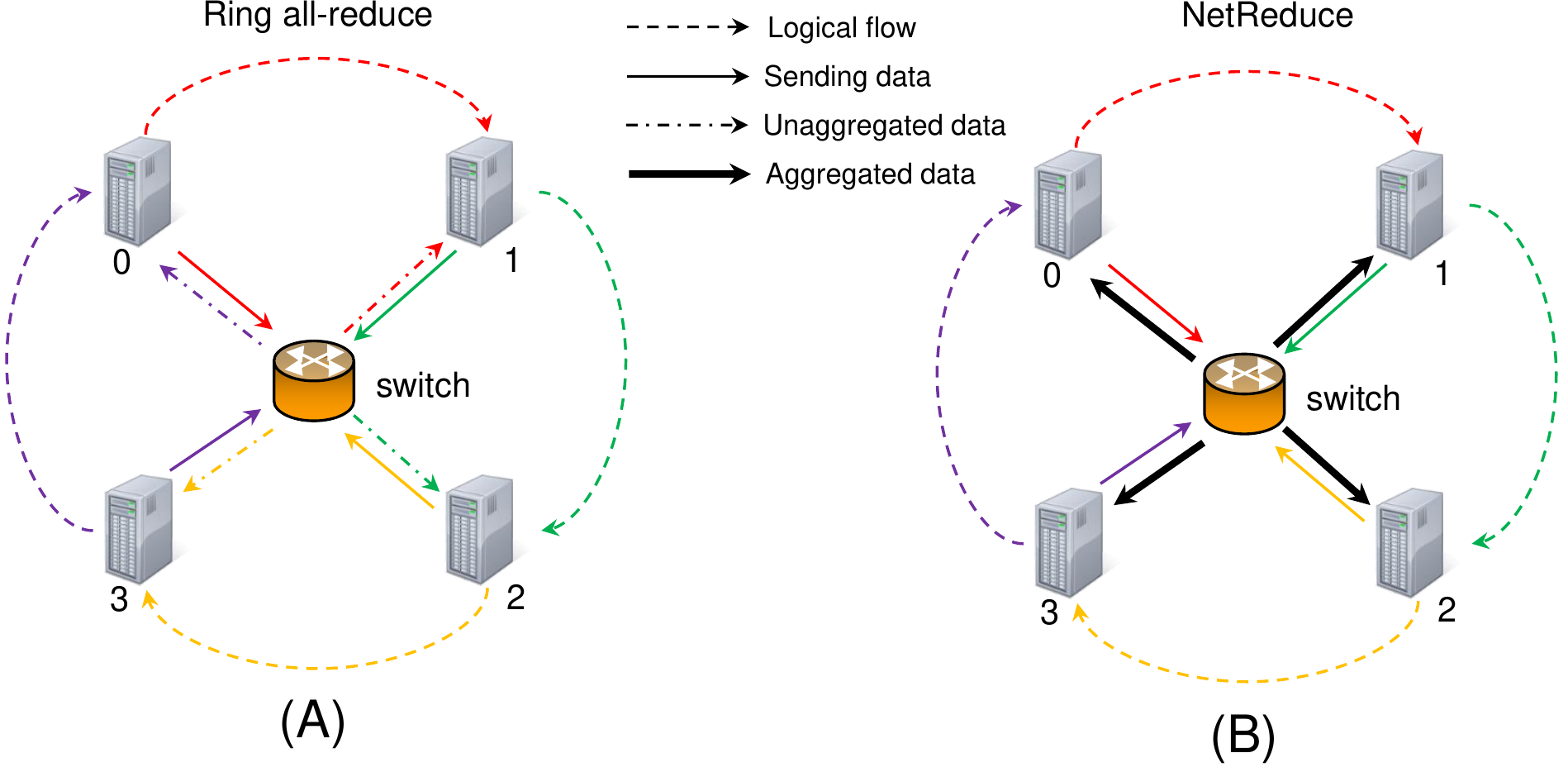}}
	\caption{Distributed DNN training strategies: (A) ring all-reduce: different nodes receive different unaggregated data from the other node; (B) NetReduce: different nodes receive the same aggregated data from all nodes.}
	\label{fig:framework}
\end{figure}

Similar to ring all-reduce, NetReduce forms a logical ring (colorful dashed curves in Figure~\ref{fig:framework}) where each node only communicates to its neighboring nodes. This is because RoCE v2 only supports a point-to-point connection and the maintenance of such a connection minimizes the number of state machines implemented in the switch. In ring all-reduce, each node receives different data (colorful dash-dotted lines in Figure~\ref{fig:framework}(A)) from the other node, which is unaggregated and the same as what its neighbor sends (colorful solid lines in Figure~\ref{fig:framework}(A)). In NetReduce, on the contrary, different nodes receive the same data, which is the aggregation result from all nodes (bold solid lines in Figure~\ref{fig:framework}(B)). When the packets from different nodes arrive at the NetReduce switch, the switch aggregates the gradients at the same neural network layer, replaces the payload of the packets with the aggregation result, and forwards them.

Another difference of NetReduce from ring all-reduce lies in the sending order of the data. In ring all-reduce, different nodes need to transmit the data in different layers to optimize bandwidth utilization. After receiving the data block, the nodes aggregate the corresponding gradients themselves. Instead, NetReduce offloads the gradients aggregation from end-hosts to network switches and ensures the aggregation correctness by letting different nodes send the same piece of data. 

%-------------------------------------------------------------------------------
\subsection{Multiple GPUs in A Machine}\label{subsec:mulgpu}
%-------------------------------------------------------------------------------

A straightforward approach in the multi-machines multi-GPUs scenario is to consider every GPU as homogeneous, regardless of the intra ones (connected via expansion bus inside a single machine) or the inter ones (connected via computer network between different machines). All GPUs are hence connected through a big flat ring but there would be a bandwidth gap between the intra ring (e.g., PCIe or NVLink) and the inter ring (e.g., Ethernet or InfiniBand). 

\begin{figure}[t]
	\centerline{\includegraphics[scale=0.4]{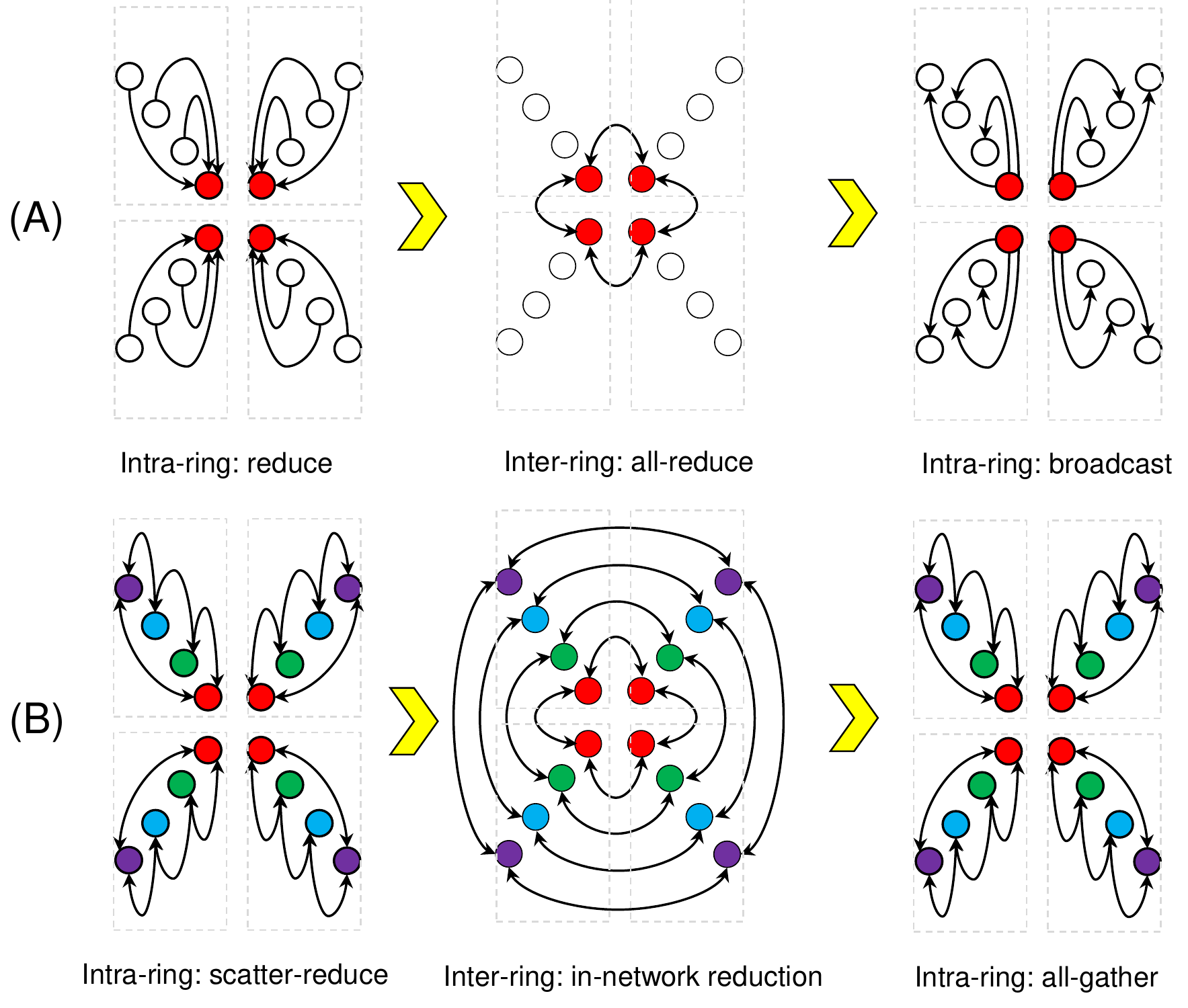}}
	\caption{Hierarchical all-reduce: (A) Tencent all-reduce: red and white circles refer to master and slave GPUs, respectively; (B) NetReduce: circles with the same color belong to the same aggregation ring. }
	\label{fig:hierarchical}
\end{figure}

Another strategy, called \textit{hierarchical all-reduce}, first aggregates parameters inside every single machine and then exchanges the aggregation results between the machines. Tencent proposed a three-phase hierarchical all-reduce algorithm in~\cite{jia2018highly}, of which the process is as shown in Figure~\ref{fig:hierarchical}(A). In the first phase, GPUs in the intra rings perform \textit{reduce} operation. Unlike that all-reduce operation writes data to every GPU, reduce operation writes the local aggregation result only to a single master GPU. In the second phase, the master GPUs among different machines form an inter ring and perform \textit{all-reduce} operation. At the moment, the master GPUs get the global aggregation result from all GPUs. Finally, each master GPU broadcasts the global aggregation result to the other local GPUs inside the same machine. In this approach, the master GPUs face a heavy burden while the computation resource of the other GPUs is wasted.

Hierarchical NetReduce, on the contrary, fully utilizes all GPUs. Suppose that a homogeneous distributed system has a total of $P$ GPUs, and each machine is equipped with $n$ ($n \geq 2$) GPUs where $P$ is an integer multiple of $n$. Therefore, the number of machines $H$ equals $P/n$. The process of hierarchical NetReduce is shown in Figure~\ref{fig:hierarchical}(B) which also consists of three phases. In the first phase, NetReduce performs \textit{scatter-reduce} operation in the intra ring, with each GPU having a partial aggregation result of a different data block (with a size of $\frac{M}{n}$) from any other one in the same machine. In the second phase, the GPUs corresponding to the same data block in different machines form multiple inter rings, performing in-network reduction simultaneously. The number of inter rings equals $n$. At the moment, each GPU has a global aggregation result on a specific data block whose size is $\frac{M}{n}$. In the third phase, NetReduce performs \textit{all-gather} operation in the intra ring that all GPUs finally obtain the global aggregation result on the data with an amount of $M$.    

In the multi-machines multi-GPUs scenario, the communication time taken by using the flat ring all-reduce algorithm is modeled as
\begin{equation}
T_{fr}=2(P-1)\alpha + 2\frac{P-1}{P}\frac{M}{B_{inter}} \label{eq:fr}
\end{equation}
where  $B_{inter}$ refers to the bandwidth in the inter ring where machines are connected via computer networks such as Ethernet or InfiniBand.

For Tencent all-reduce, consider Rabenseifner's reduce algorithm~\cite{rabenseifner1997new} and Van de Geijn's broadcast algorithm~\cite{barnett1994interprocessor}, and assume $n$ is a power of 2, the communication cost can be modeled as
\begin{eqnarray}
T_{tr} &=& T_{tr1} + T_{tr2} +T_{tr3} \notag \\
&=& \left[2\alpha\log_{2}(n) + \frac{2(n-1)}{n}\frac{M}{nB_{intra}}\right] \notag \\
&& + \left[2(\frac{P}{n}-1)\alpha + 2\frac{P/n-1}{P/n}\frac{M}{B_{inter}}\right] \notag \\
&& + \left[(\log_{2}(n)+n-1)\alpha + 2\frac{n-1}{n}\frac{M}{B_{intra}}\right] \notag \\
&=&   \frac{n^2+3n\log_{2}(n)-3n+2P}{n}\alpha \notag \\ && + \frac{4(n-1)PB_{inter}+2(P-n)nB_{intra}}{nPB_{intra}B_{inter}}M 
\label{eq:tr}
\end{eqnarray}
where  $B_{intra}$ refers to the bandwidth of the intra ring where GPUs are connected via expansion bus such as PCIe or NVLinks.

The communication cost of hierarchical NetReduce is given as
\begin{eqnarray}
T_{nh} &=& T_{nh1} + T_{nh2} +T_{nh3} \notag \\
&=& \left[(n-1)\alpha + (n-1)\frac{M}{nB_{intra}}\right] + \left(\alpha + \frac{M}{B_{inter}}\right) \notag \\
&& + \left[(n-1)\alpha + (n-1)\frac{M}{nB_{intra}}\right] \notag \\
&=&   (2n-1)\alpha + \frac{2(n-1)B_{inter}+nB_{intra}}{nB_{intra}B_{inter}}M 
\label{eq:nh}
\end{eqnarray}
When $n=1$, $B_{intra}=B_{inter}=B$, Eq.(\ref{eq:nh}) reduces to Eq.(\ref{eq:netreduce}).

Eq.(\ref{eq:tr}) subtracting Eq.(\ref{eq:nh}) gives 
\begin{eqnarray}
\Delta T_{tr-nh} &=& T_{tr} - T_{nh} \notag \\
&=& (2P/n+3\log_{2}(n)-n-2)\alpha \notag \\
&& + \frac{(P-2n)nB_{intra}+2(n-1)PB_{inter}}{nPB_{intra}B_{inter}}M
\label{eq:deltatr}
\end{eqnarray}
When $P > 3n$, (\ref{eq:deltatr}) is always larger than 0, considering $n$ is usually no larger than 16.

Eq.(\ref{eq:fr}) subtracting Eq.(\ref{eq:nh}) gives 
\begin{eqnarray}
\Delta T_{fr-nh} &=& T_{fr} - T_{nh} \notag \\
&=& (2P-2n-1)\alpha \notag \\
&& + \frac{(P-2)nB_{intra}-2(n-1)PB_{inter}}{nPB_{intra}B_{inter}}M
\label{eq:deltafr}
\end{eqnarray}

Similarly, we can obtain a relaxed sufficient condition from (\ref{eq:deltafr}) that hierarchical NetReduce outperforms flat ring all-reduce on communication as follows
\begin{equation}
\frac{B_{intra}}{B_{inter}} \geq \frac{2P}{P-2} \qquad (P > n \geq 2)
\label{eq:condition}
\end{equation}

%-------------------------------------------------------------------------------
\section{NetReduce Design}\label{sec:design}
%-------------------------------------------------------------------------------

%-------------------------------------------------------------------------------
\subsection{Network Protocol}\label{subsec:protocol}
%-------------------------------------------------------------------------------
A NetReduce protocol is designed to enable the coordination between the end-host and the network switch. The new protocol can be regarded as an L4.5 protocol embedded in the L4 payload. NetReduce uses existing L2/L3 routing protocols to forward packets.

\begin{figure}[t]
	\centerline{\includegraphics[scale=0.38]{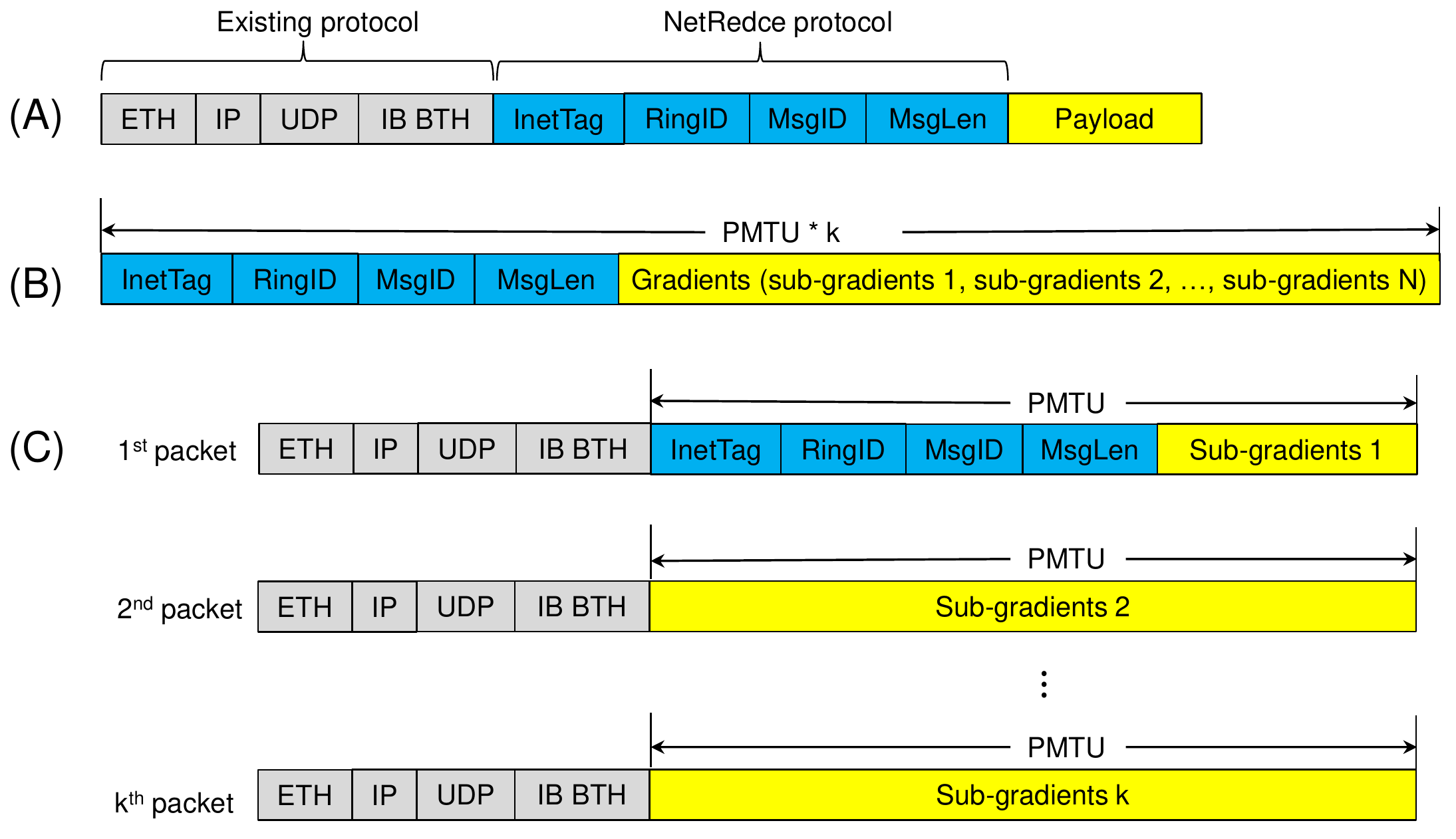}}
	\caption{NetReduce protocol: (A) NetReduce header format; (B) an RDMA data message; (C) corresponding packets of the message after segmentation.}
	\label{fig:protocol}
\end{figure}

The NetReduce protocol is placed after the IB Base Transport Header (BTH), and the header format is shown in Figure~\ref{fig:protocol}(A). The major fields consist of \textit{InetTag}, \textit{RingID}, \textit{MsgID}, \textit{MsgLen}. \textit{InetTag} marks the aggregation packet. \textit{RingID} indicates in which specific ring that the current packet participates. For example, there are multiple rings between different machines in the multi-machines multi-GPUs scenario discussed in~\cref{subsec:mulgpu}. Only the message with the same \textit{MsgID} in the same ring from different computing nodes can be aggregated since they contain the same gradient variables at the same neural network layer. \textit{MsgLen} denotes the number of packets per message after NIC segmentation. 

Note that not every aggregation packet contains such a NetReduce header. The NetReduce header is inserted at the beginning of each RDMA message as shown in Figure~\ref{fig:protocol}(B). To comply with the Maximum Transmission Unit (MTU) in the Ethernet link layer, a to-be-sent message is segmented into multiple packets by NIC according to the Path MTU (PMTU). Suppose that the size of the message is PMTU$*k$, the message will then be segmented to $k$ packets as shown in Figure~\ref{fig:protocol}(C). Among all the $k$ packets in the message, the NetReduce header only appears in the first one. We do not modify the segmentation behavior of NIC. For those packets without the NetReduce header, we recover their corresponding NetReduce information which will be described in detail in~\cref{subsec:accelerator}.         

%-------------------------------------------------------------------------------
\subsection{Message-Level Flow Control}\label{subsec:flowcontrol}
%-------------------------------------------------------------------------------

To prevent buffer overflow, SwitchML sends messages one-by-one, i.e., keeping only one message unacknowledged and sending the next message after receiving the acknowledgment. However, this stop-and-wait transmission makes it difficult to achieve full bandwidth utilization~\cite{peng2019generic}. NetReduce introduces a message-level credit-based flow control mechanism by using a sliding window. The credit is the aggregation result of the previous message. The basic idea is that end-hosts first send $N$ (refers to the window size) pieces of message concurrently to fully utilize bandwidth. They will not send the $(N+i)^{th}$ message until the aggregation result of the $i^{th}$ message is received. For example, if $N$=3, end-hosts do not send the $4^{th}$ messages until the aggregation result of the $1^{st}$ messages is received. Similarly, they can only send the $5^{th}$ messages until they receive the aggregation result of the $2^{nd}$ messages and so on.

Algorithm~\ref{algo:worker} describes the processing algorithm of end-hosts which deals with the message-level data. Basically, the ``send-receive'' processes appear \textit{NumMsg}-times, i.e., every time a nodes sends out a message it would expect to receive the corresponding aggregation result from all nodes on this message. For each sending, a message, \textit{msg}, will be assigned a NetReduce header, including \textit{InetTag}, \textit{RingID}, \textit{MsgID} and \textit{MsgLen}. The window size, $N$, can be theoretically calculated based on the buffer size that the switch is able to provide as follows
\begin{eqnarray}
N \times MsgLen \times pktSize &\geq& RTT \times PortRate \notag \\
N &\geq& \frac{RTT \times PortRate}{MsgLen \times pktSize} 
\label{eq:sw}
\end{eqnarray} 
where $RTT$ and $PortRate$ refer to the time taken by completing the transmission of a single packet in the system and the bandwidth of NIC bandwidth at the end-host, respectively. 
\begin{algorithm}[t]
	\caption{The end-host processing algorithm.}
	\label{algo:worker}
	\begin{algorithmic}[1]
		\REQUIRE ~~\\
		Total number of messages to be transmitted, \textit{NumMsg}; \\
		Number of messages per sliding window, $N$; \\
		\IF{\textit{NumMsg} $\leq$ $N$}
		\STATE $N$ = \textit{NumMsg};
		\ENDIF
		\STATE /* Send the first $N$ messages */
		\FOR{i in 0 : ($N$ - 1)}
		\STATE msg.\textit{InetTag} = \textit{InetTag}; 
		\STATE msg.\textit{RingID} = \textit{RingID}; msg.\textit{MsgID} = \textit{MsgID};
		\STATE msg.\textit{MsgLen} = \textit{MsgLen};
		\STATE msg.\textit{Params} = Tensor[\textit{RingID}][\textit{MsgID}];
		\STATE send(msg);
		\STATE \textit{MsgID} ++;
		\ENDFOR
		\STATE /*Send $(N+i)^{th}$ message after $i^{th}$ message is received*/
		\FOR{i in $N$ : (\textit{NumMsg} - 1)}
		\STATE receive(msg);
		\STATE Tensor[msg.\textit{RingID}][msg.\textit{MsgID}] = msg.\textit{Params};
		\STATE msg.\textit{InetTag} = \textit{InetTag}; msg.\textit{RingID} = \textit{RingID};
		\STATE msg.\textit{MsgID} = msg.\textit{MsgID} + $N$;
		\STATE msg.\textit{MsgLen} = \textit{MsgLen};
		\STATE msg.\textit{Params} = Tensor[msg.\textit{RingID}][msg.\textit{MsgID}];
		\STATE send(msg);
		\ENDFOR
		\STATE /* Receive the last $N$ messages */
		\FOR{i in 0 : ($N$ - 1)}
		\STATE receive(msg);
		\STATE Tensor[msg.\textit{RingID}][msg.\textit{MsgID}] = msg.\textit{Params};
		\ENDFOR
	\end{algorithmic}
\end{algorithm}

%-------------------------------------------------------------------------------
\subsection{In-Network Reduction Accelerator}\label{subsec:accelerator}
%-------------------------------------------------------------------------------

\begin{figure*}[t]
	\centerline{\includegraphics[scale=0.7]{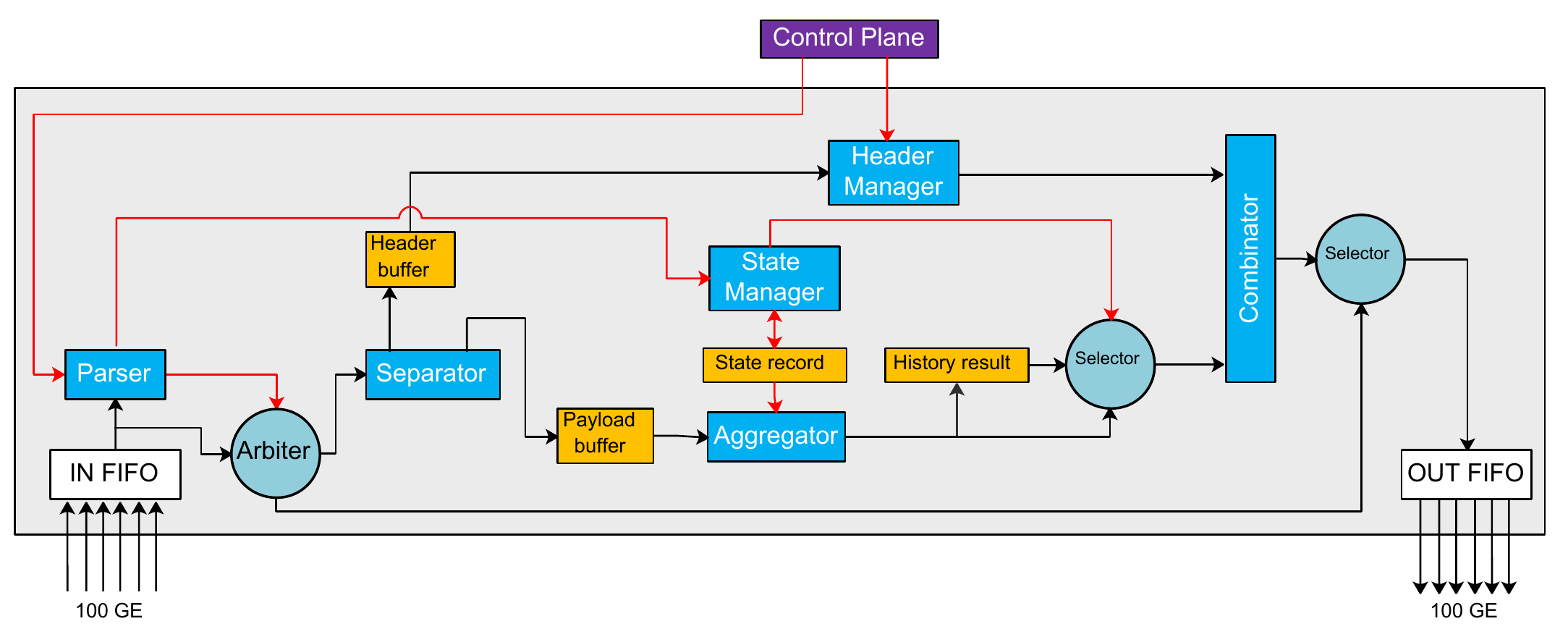}}
	\caption{The accelerator architecture of in-network reduction (red and black arrow lines refer to control and data flows, respectively).}
	\label{fig:fpga}
\end{figure*}
The in-network reduction accelerator is designed as a middlebox attached to the Ethernet switch which we do not modify. We describe the accelerator architecture in Figure~\ref{fig:fpga}. When a packet arrives, a \texttt{Parser} identifies the aggregation packet or directs the other kinds of the packet to the output port directly. The \texttt{Parser} further feeds the NetReduce header to a \texttt{State Manager} which tracks the arrival states of the packets. 

A \texttt{Separator} will separate the protocol headers (including Ethernet, IP, UDP, BTH, and NetReduce) from the payload. The headers are then fed to a \texttt{Header Manager} which decides on single-switch aggregation or multi-switches aggregation (See~\cref{subsec:clos}). Once the arrival states of packets from all workers are valid, the \texttt{Aggregator} begins to sum the payload at the granularity of packets and writes the aggregation result to a buffer which stores the history results. The \texttt{Combinator} finally merges original headers and updated payload to one complete packet, and sends it out. 
%NetReduce only updates the payload and does not change the packet header.

\subsubsection{Header Recovery for Non-First Packets}\label{subsubsec:recovery} 

As mentioned in \cref{subsec:protocol}, only the first packet in a message contains the NetReduce header due to the NIC segmentation. We propose an algorithm to recover the header information for those non-first packets. NetReduce uses a three-element tuple \{\textit{SrcIP}, \textit{DstIP}, \textit{DstQP}\} to uniquely determine an RDMA connection where \textit{SrcIP} (Source IP Address) and \textit{DstIP} (Destination IP Address) are in the IP header, \textit{DstQP} (Destination Queue Pair) is in the IB BTH. Since the packets in the same message belong to one single RDMA connection, the non-first packets are thus connected to their first packets via the tuple.

\begin{figure}[t]
	\centerline{\includegraphics[scale=0.5]{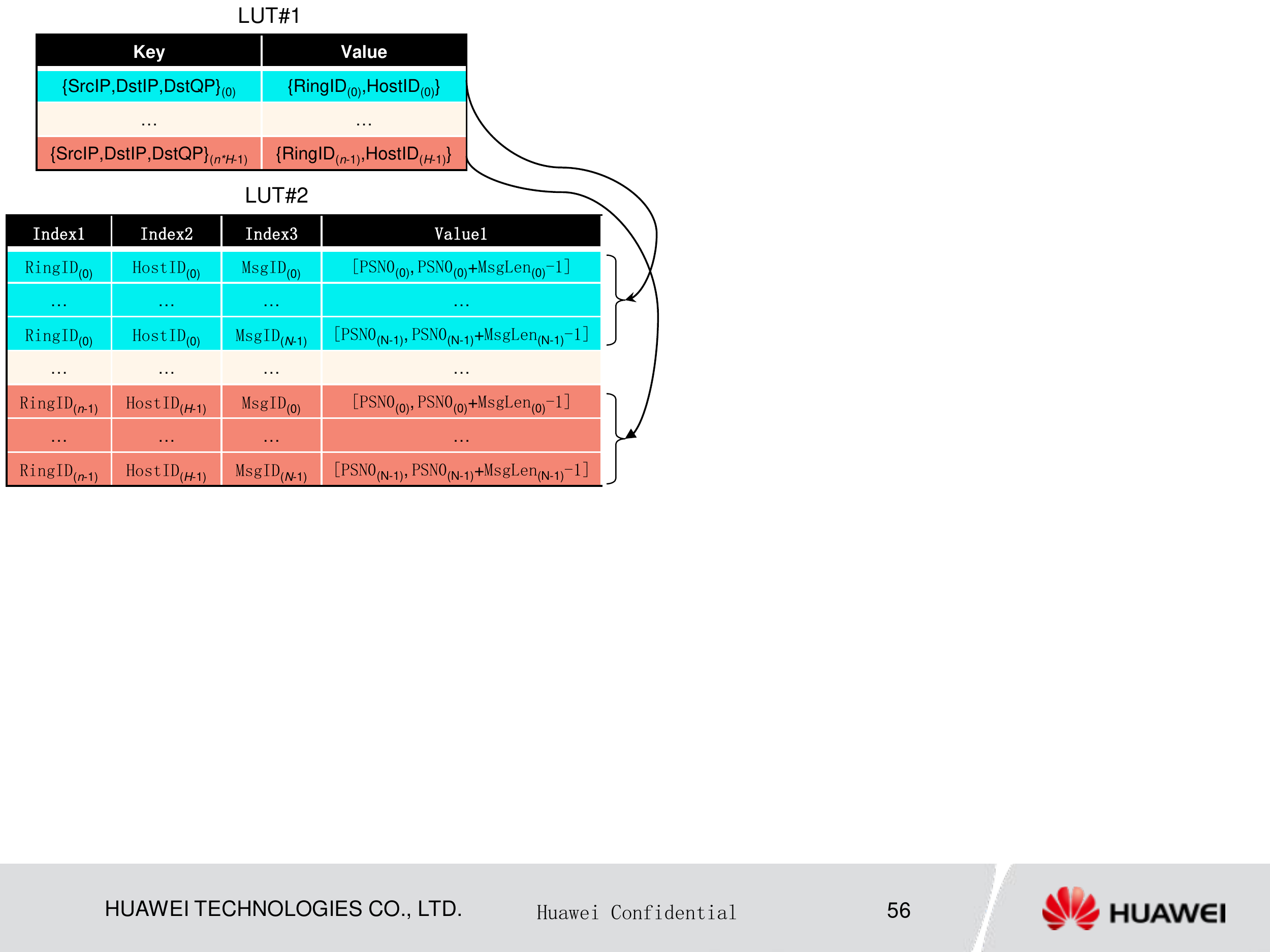}}
	\caption{A two-level lookup table to recover ring (LUT\#1) and message (LUT\#2) information, respectively.}
	\label{fig:lut}
\end{figure}

The \texttt{Parser} maintains a two-level LUT (lookup table) as shown in Figure~\ref{fig:lut}. LUT\#1 and LUT\#2 recovers the ring and the message information, respectively. When a first packet marked by \textit{InetTag} arrives, the \texttt{Parser} locates the end-host in the ring (identified by \textit{RingID}) by assigning a \textit{HostID} to the three-element tuple of the packet. Note that the \textit{RingID} is assigned by end-hosts but the \textit{HostID} is counted by the switch. Then LUT\#1 records the tuple and the corresponding \textit{RingID} and \textit{HostID}. The number of entries in LUT\#1 is $n*H$, where $n$ is the number of GPUs in a single machine (equals the number of rings performing all-reduce) and $H$ refers to the number of machines per ring. Since one tuple-defined RDMA connection corresponds to only one ring, the \textit{RingID} can be recovered solely by the tuple.

Since an RDMA connection may consist of multiple pieces of message, solely using the tuple cannot recover \textit{MsgID}. To map a packet into a specific message, NetReduce uses Packet Sequence Number (PSN) in the IB BTH and \textit{MsgLen} in the NetReduce protocol. Suppose the PSN of a first packet in a message is \textit{PSN0}, then the PSN of packets with the same tuple falling in the range of [\textit{PSN0, PSN0 + MsgLen - 1}] belongs to the same message. LUT\#2 records the \textit{PSN} and \textit{MsgLen} of the first packets and recovers the \textit{MsgID} for non-first packets by locating the entry where the \textit{PSN} belongs. The number of the LUT\#2 entries is $n*H*N$, where $N$ refers to the sliding window size (number of messages per window). The values of $n$, $H$, and $N$ are determined via the control plane at the job initialization period.
\begin{algorithm}[t]
	\caption{Recovery algorithm of NetReduce header.}
	\label{algo:mapping}
	\begin{algorithmic}[1]
		\IF{the packet is marked by $InetTag$}
		\STATE Assign $HostID$ to [$SrcIP$,$DstIP$,$DstQP$];
		\STATE Create entries in LUT\#1 and LUT\#2, respectively;
		\ELSE
		\IF{[$SrcIP$,$DstIP$,$DstQP$] has been recorded in LUT\#1 previously}
		\STATE Recover $RingID$ and $HostID$ from LUT\#1;
		\STATE Recover $MsgID$ from LUT\#2 by looking up which range [\textit{PSN0, PSN0 + MsgLen - 1}] the current $PSN$ belongs to;
		\ELSE
		\STATE Direct the packet to output port;
		\ENDIF   
		\ENDIF
	\end{algorithmic}
\end{algorithm}

The recovery algorithm works in a lossy network. The RDMA RC mode guarantees strictly ordered transmission. If the first packet is lost or out-of-ordered, the sender retransmits the whole message. The receiver does not receive the packets belonging to the same RDMA connection until the previous lost first packet successfully arrives. The whole recovery algorithm is summarized in Algorithm ~\ref{algo:mapping}.

\subsubsection{Aggregation and Dealing with Packet Loss}\label{subsubsec:pktloss} 

\begin{figure}[t]
	\centerline{\includegraphics[scale=0.8]{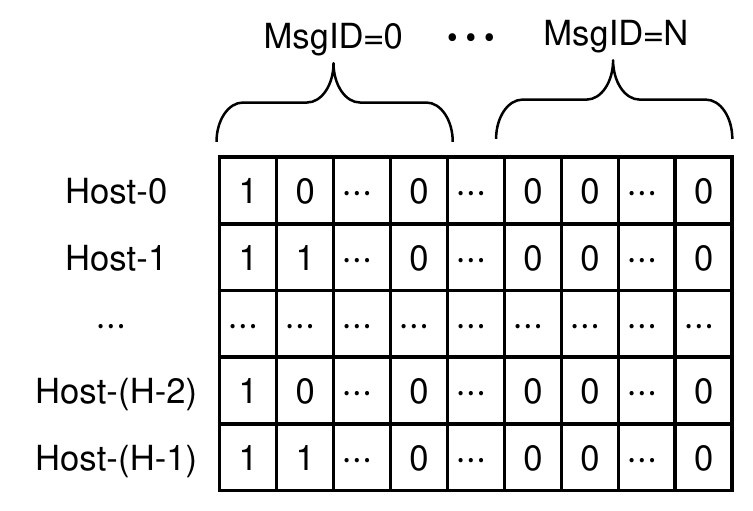}}
	\caption{Arrival states of packets for each ring.}
	\label{fig:bitmap}
\end{figure}

The accelerator does not start to aggregate gradients until the corresponding packets from all end-hosts arrive. The aggregation packets are aligned with the same \textit{RingID}, \textit{MsgID} and PSN offset from \textit{PSN0}. The \texttt{State Manager} assigns a bit to each packet in the sliding window to indicate the arrival state of the packet, resulting in a bitmap for each ring shown in Figure~\ref{fig:bitmap}. The bitmap contains packet states in ($N+1$) pieces of message. When an aggregation packet arrives, the \texttt{State Manager} first locates the specific bitmap based on $RingID$ and then set the state according to the index [$HostID$, $PSN-PSN0+(MsgID+1)\%(N+1)-1$] in the matrix. Once all the elements in a column equal 1, the \texttt{Aggregator} aggregates the corresponding gradients and writes the aggregation result to a history buffer.

When the aggregation completes, the states are not set to 0 immediately. The states of the $i^{th}$ message cannot be updated until the aggregation results of the message are received by end-hosts (i.e., when the $(N+i)^{th}$ message arrives). Specifically, the \texttt{State Manager} not only sets the state at [$HostID$, $PSN-PSN0+(MsgID+1)\%(N+1)-1$] to 1 but also updates the state at [$HostID$, $PSN-PSN0+(MsgID+1)\%(N+1)$] to 0. This is to ensure that a packet state will not be updated before its aggregation result is successfully received by end-hosts. Then the \texttt{State Manager} can identify retransmitted packets by checking the corresponding arrival states in the bitmap. If the value of the state is 1 already, the \texttt{State Manager} knows it is a retransmitted packet. Then the \texttt{State Manager} makes a decision depending on whether this packet has been aggregated previously (i.e. if all elements in the corresponding column equal to 1). If yes, the accelerator replaces the packet payload with the aggregation result in history record and directs it to the output port. On the other hand, if the parameter has not been aggregated, the accelerator simply discards the packet. 

%-------------------------------------------------------------------------------
\subsection{Implementation}\label{subsec:implementation}
%-------------------------------------------------------------------------------

\begin{figure}[t]
	\centerline{\includegraphics[scale=0.7]{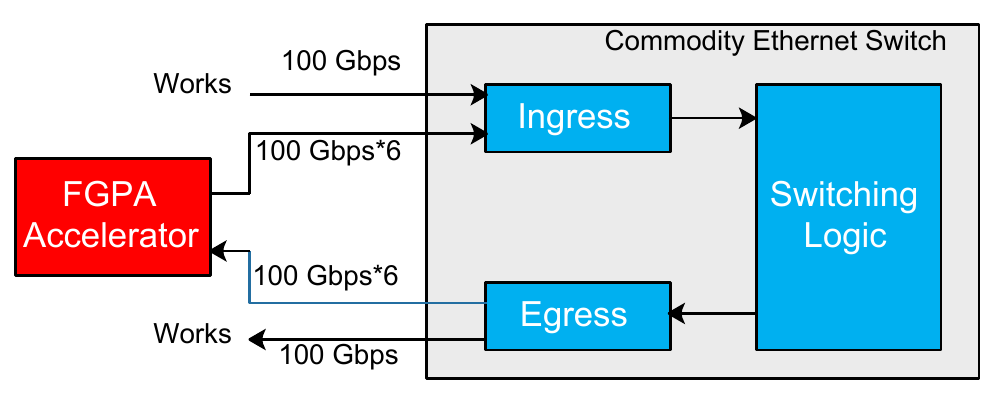}}
	\caption{NetReduce hardware prototype.}
	\label{fig:networkdevice}
\end{figure}

We develop a prototype by using a 100 GbE commodity switch for basic forwarding as shown in Figure~\ref{fig:networkdevice}. A self-developed FPGA board with NetReduce capability is attached to the switch. The FPGA board is equipped with a Xilinx Virtex Ultrascale chip~\cite{xilinx} which is able to support at most 100 Gbps $\times$ 6. 

We configure the switch ACL rules to redirect all RoCE v2 packets to FPGA and forward other packets directly. The FPGA further differentiates the aggregation packets (including the first and non-first packets in an RDMA message) from the other RoCE v2 packets if there exists any. In this way, we do not need to augment the switch capability and a commodity switch would work. The FPGA then processes aggregation packets with the NetReduce logic and send the packets with aggregation results back to the switch. The switch then forwards those packets to proper end-hosts by using existing L2/L3 routing protocols. The switch needs to process the packets twice, thus increasing the latency. Nevertheless, our evaluation shows the additional FPGA operations add less than 3 $\mu$s extra RTT compared with that the original RTT is 2 $\mu$s. 

%-------------------------------------------------------------------------------
\subsection{Extension to Spine-Leaf Topology }\label{subsec:clos}
\begin{figure}[t]
	\centerline{\includegraphics[scale=0.5]{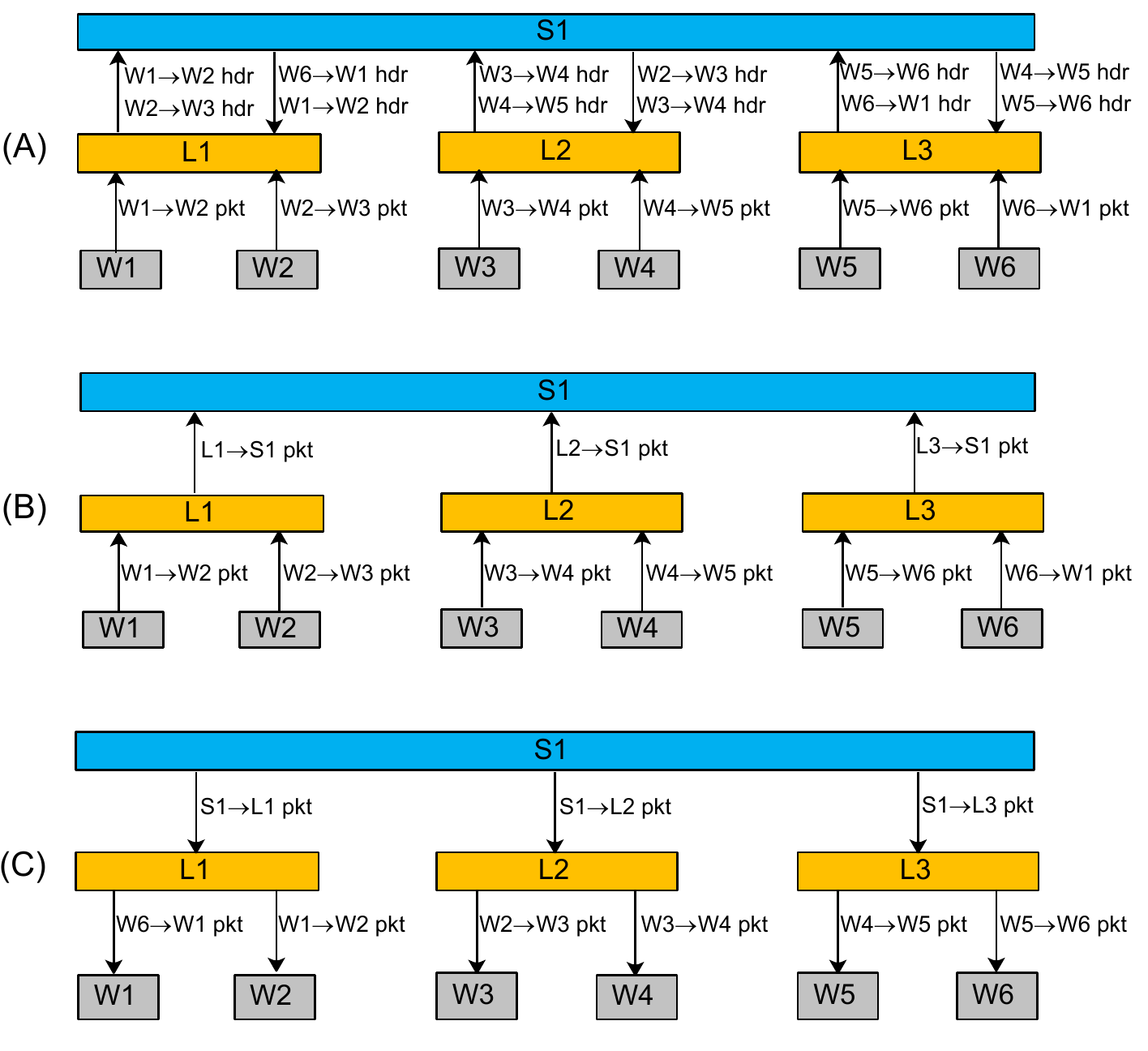}}
	\caption{Data flow in spine-leaf topology. Upstream: (A) leaves send headers of aggregation packets to destination leaves; (B) leaves send packets with modified headers and payload to the spine. Downstream: (C) leaves combine original headers and aggregation results and send to workers.}
	\label{fig:address}
\end{figure}

%-------------------------------------------------------------------------------
In this subsection, we extended the in-network operations from the rack-level cluster to a more general spine-leaf topology as shown in Figure~\ref{fig:address}. Figure~\ref{fig:address} shows an example where packets from 6 workers are aggregated via two-level switches. The major difference in this topology from the rack-scale aggregation is that NetReduce modifies not only the payload but also the header. 

During the job initialization period, an aggregation tree is formed by binding a spine to the leaves. For example, a spine can be selected with the smallest value of IP address. The control plane informs the values of two state variables to the \texttt{Header Manager} (Figure~\ref{fig:fpga}) in the leaf: $LocalSize$ and $GlobalSize$, which refer to the numbers of local machines under the leaf and global machines in the whole training job, respectively. Specifically, there are two different settings: $LocalSize=H=GlobalSize$ and $LocalSize=H<GlobalSize$ for only Top-on-Rack (ToR) switch aggregation and two-level switches aggregation, respectively. The operation process of the \texttt{Header Manager} is described in Algorithm~\ref{algo:address}. 
\begin{algorithm}[t]
	\caption{Processing algorithm of \texttt{Header Manager}.}
	\label{algo:address}
	\begin{algorithmic}[1]
		\IF{$LocalSize$ == $GlobalSize$}
		\STATE Does not change the packet headers;
		\ELSIF{$LocalSize$ < $GlobalSize$}
		\STATE Send the packet headers to the destination leaf;
		\STATE Change the original [$SrcMAC,DstMAC,SrcIP,DstIP$] to [$SrcMAC_{leaf},DstMAC_{spine},SrcIP_{leaf},DstIP_{spine}$];
		\ENDIF
		\IF{[$DstMAC,DstIP$] belongs to the switch itself}
		\STATE For spine, swap [$SrcMAC_{leaf},SrcIP_{leaf}$] and [$DstMAC_{spine},DstIP_{spine}$];
		\STATE  For leaf, replace the headers with the ones previously stored based on $DstQP$ and $PSN$;
		\ENDIF
	\end{algorithmic}
\end{algorithm}

Lines 1 to 5 in Algorithm~\ref{algo:address} apply to the leaf switch only, which differentiate the ToR aggregation and the two-switches aggregation. For the latter case, in the upstream flow, the leaves first send the packet headers to the destination leaf as shown in Figure~\ref{fig:address}(A). L1, L2, and L3 stores the headers of (W6$\to$W1,W1$\to$W2), (W2$\to$W3,W3$\to$W4),(W4$\to$W5,W5$\to$W6), respectively. These headers are used by the leaves to distribute packets back to workers in the downstream flow. Then the leaves send the local aggregation results to the spine with source and destination addresses replaced by the ones of the leaves themselves and the spine, respectively, as shown in Figure~\ref{fig:address}(B). In the downstream flow as shown in Figure~\ref{fig:address}(C), the spine swaps source and destination addresses and sends the packets with global aggregation results to the leaves. The leaves replace the headers with the previously stored ones and send the whole packets to the workers. These actions of changing headers are triggered by detecting whether the destination address belongs to the switch itself.

%-------------------------------------------------------------------------------
\section{Evaluation}\label{sec:evaluation}
%-------------------------------------------------------------------------------

\subsection{Methodology}

We compare the proposed NetReduce with ring all-reduce and SwitchML. The ring all-reduce is implemented by NCCL-2.4.7~\cite{nvidia2019nccl}, a commonly used collective communication library for distributed DNN training, while SwitchML is implemented by using a programming switch equipped with a Tofino chip. We modify the primitive in NCCL-2.4.7 and create a new \textit{GenericOp} to comply with the NetReduce function. 

In multi-machines single-GPU scenario, we use 6 servers. Each server is equipped with two 10-cores CPUs (Intel Xeon E5-2064 2.4 GHz), 32 GB*3 DDR4 memory, one NVIDIA Geforce RTX 2080 8 GB GPU~\cite{nividia2080}, and a Mellanox ConnectX-5~\cite{mellanoxcx5} 100 GbE NIC. For multi-machines multi-GPUs, we use 4 servers. Each server is equipped with two 18-cores CPUs (Intel Xeon Gold 6154 3.00 GHz), 1 TB (64 GB*16) DDR4 memory, eight NVIDIA Tesla V100 SXM2 32 GB GPUs~\cite{nvidia2020v100} and a Mellanox ConnectX-5 100 GbE NIC. A hybrid cube-mesh network topology~\cite{nvidia2017dgx} is used for 8-GPU interconnection via NVLink~\cite{nvidia2019nvlink} inside each single machine.   

We evaluate the systems in typical image training workload, ImageNet~\cite{deng2009imagenet}. Three representative Convolutional Neural Network (CNN) models are chosen: AlexNet~\cite{krizhevsky2012imagenet}, VGG-16~\cite{simonyan2014very}, and ResNet-50~\cite{he2016deep}. We leverage Horovod-0.16.0~\cite{sergeev2018horovod} to support TensorFlow-1.12.0~\cite{tensorflow2019benchmark}. We also evaluate NetReduce on NLP tasks by using PyTorch-1.5.1~\cite{pytorch}. We pretrains transformer-based models (BERT\cite{devlin2018bert} and GTP~\cite{radford2019better}) and fine-tunes the model for GLUE~\cite{wang2018glue} and SQuAD~\cite{rajpurkar2016squad} tasks by using the approach provided in~\cite{shoeybi2019megatron} and~\cite{transformer2020}, respectively. In the experiments, the sliding window size $N=2$, the message size is 170 KB, and each packet delivers 1 KB of payload data.

%-------------------------------------------------------------------------------
\subsection{Multi-Machines Single-GPU Scenario}\label{subsec:eval_mmsg}
%-------------------------------------------------------------------------------

\begin{figure}[t]
	\centerline{\includegraphics[scale=0.5]{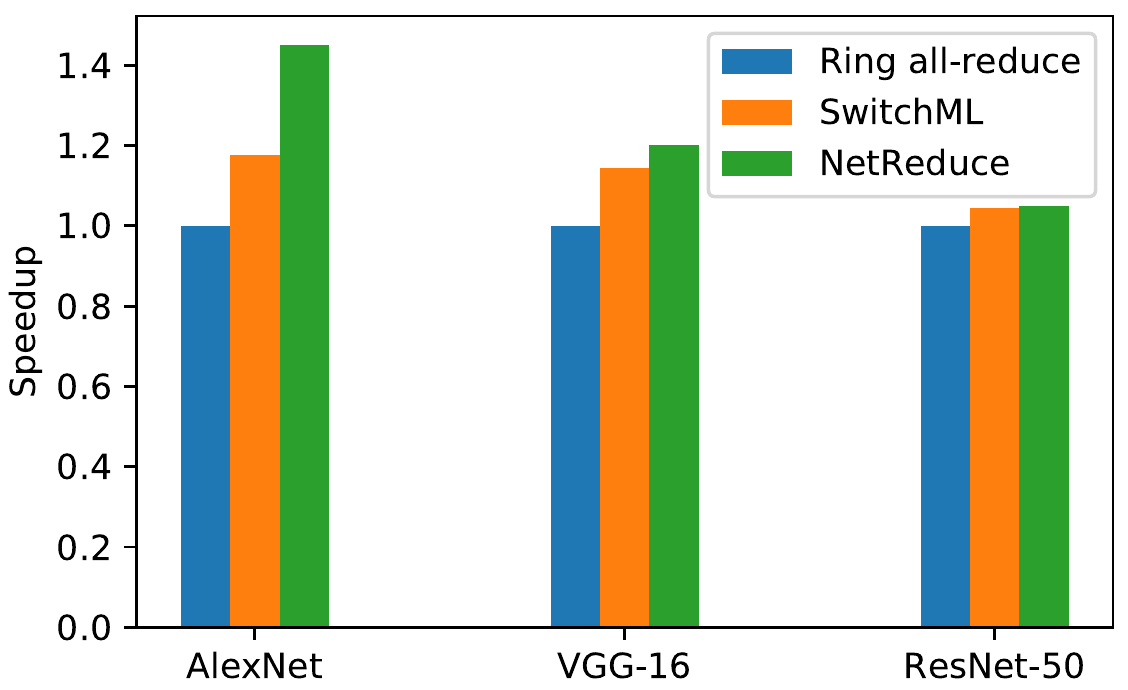}}
	\caption{Speedup of in-network aggregation over ring all-reduce on CNN models by using 6 NVIDIA GeForce RTX 2080 GPUs.}
	\label{fig:switchml}
\end{figure}

We plot the speedup of in-network aggregation over ring all-reduce by using 6 GeForce RTX 2080 GPUs in Figure~\ref{fig:switchml}, where the image training throughput of SwitchML and NetReduce is normalized to that of ring all-reduce. SwitchML improves the baseline on the three models by 17.5\%, 14.4\%, and 4.4\% respectively while the percentages by using NetReduce are 45.0\%, 20.2\%, and 4.9\%, respectively. The larger performance gain of NetReduce than SwitchML mainly comes from two points: processing full-length Ethernet frame by using FGPA and offloading the network stack processing to RDMA NIC. However, NetReduce and SwitchML have similar performance on ResNet-50. This is due to that the end-to-end throughput depends on how much computation and communication overlap. In computation-intensive models such as ResNet-50, computation accounts for the most training time, and the benefits from in-network aggregation by reducing the communication time are hidden. The gain can be enlarged by decreasing the computation proportion, e.g., decreasing the batch size or training with a half-precision floating point. We will further explore these strategies.

\begin{figure}[t]
	\centerline{\includegraphics[scale=0.75]{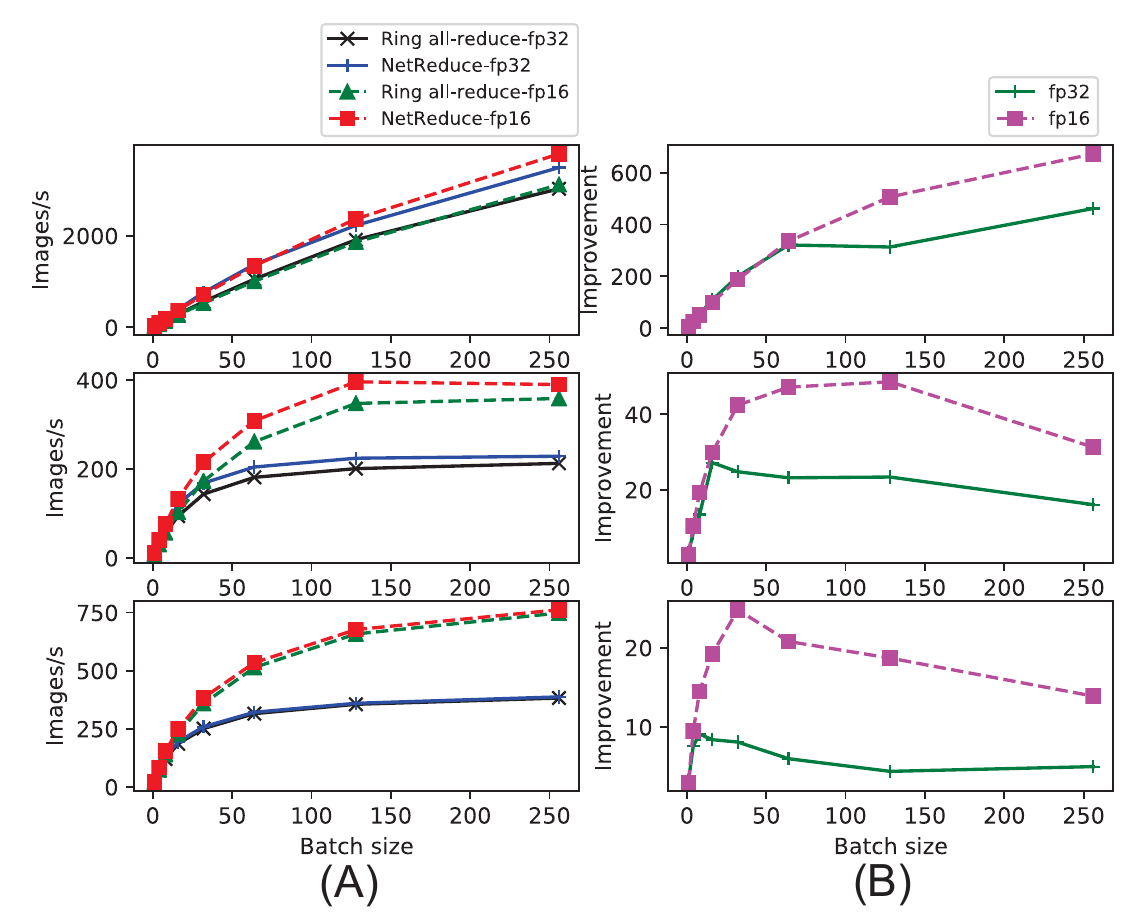}}
	\caption{Training throughput per GPU with various batch sizes and precision in multi-machines single-GPU scenario (4 NVIDIA Tesla V100): (A) image processing throughput; (B) absolute throughput improvement. The $1^{st}$ to $3^{rd}$ rows refer to AlexNet, VGG-16, and ResNet-50, respectively.}
	\label{fig:mmsg}
\end{figure}

\textbf{Effect of batch size and floating-point precision.} Since RTX 2080 has only 8 GB memory which limits the choice of batch size (BS), we change to use 4 Tesla V100 GPUs. Compared to the consumer-level RTX 2080, Tesla V100 is dedicated to AI applications in DCN with higher computing capability and much more memory (32 GB). Figure~\ref{fig:mmsg} shows the training throughput per GPU with various BS=1, 4, 8, 16, 32, 64, 128, and 256, respectively. We consider both FP32 (single precision, solid lines) and FP16 (half precision, dashed lines). 

As shown in Figure~\ref{fig:mmsg}(A), NetReduce always trains images faster than ring all-reduce for both FP32 and FP16 cases. When increasing the BS, the absolute improvement of throughput first increases and then decreases after passing some certain thresholds in the models as shown in Figure~\ref{fig:mmsg}(B), except AlexNet. This is because AlexNet is a communication-intensive model and has less requirement of GPU memory than the other models. For AlexNet, GPU can consume more data which means the ``up-down'' phenomenon would occur beyond BS=256.

In Figure~\ref{fig:mmsg}(B), using FP16 gives larger absolute improvement than FP32. This is because FP16 takes less computation time than FP32, reducing the overlap between communication and computation. Therefore, the benefits brought by NetReduce based on the reduction of communication time becomes more obvious. Take BS=32 with FP16 as an example and summarize the training performance in Table~\ref{tab:fpga}. Among the models, NetRedcue improves AlexNet on the throughput by 35.6$\%$, which is the most. This is because when using ring all-reduce, the communication accounts for 77.7$\%$ (=47.12$/$60.62 as shown in the 5$^{th}$ column in Table~\ref{tab:fpga}) of the whole iteration time, which has a significant potential to improve. On the contrary, although VGG-16 is improved on communication by 33.3$\%$ which is similar to AlexNet (34.0$\%$), the communication accounts for 60.5$\%$ which is smaller than AlexNet, resulting in a less improvement on throughput (24.5$\%$). For ResNet-50 which is a computation-intensive model, with 16.3$\%$ improvement on the communication which accounts for only 25.8$\%$ of the iteration time, we have 6.9$\%$ improvement on the throughput.
\begin{table}[t]
	\centering
	\caption{Training performance per GPU with BS=32, FP16 by using 4 NVIDIA Tesla V100.}
	\label{tab:fpga}
	\resizebox{\linewidth}{!}{
		\begin{tabular}{cc|cccc}
			\hline
			\multicolumn{2}{c|}{\multirow{2}{*}{Model}} & Throughput & Iteration & Communication \\
			& & (images/s) & (ms) & (ms) \\ \hline
			\multirow{2}{*}{AlexNet} & Ring all-reduce & 527.9 & 60.62 & 47.12(77.7\%) \\
			\multirow{2}{*}{(236 MB)} & NetReduce & 716.0 & 44.69 & 31.10(69.6\%) \\
			& $\uparrow$ & 35.6\% & 26.3\% & 34.0\% \\ \hline
			\multirow{2}{*}{VGG-16} & Ring all-reduce & 172.9 & 185.08 & 111.98(60.5\%) \\
			\multirow{2}{*}{(528 MB)} & NetReduce & 215.3 & 148.63 & 74.64(50.2\%) \\
			& $\uparrow$ & 24.5\% & 19.7\% & 33.3\% \\ \hline
			\multirow{2}{*}{ResNet-50} & Ring all-reduce & 358.8 & 89.19 & 23.04(25.8\%) \\
			\multirow{2}{*}{(98 MB)} & NetReduce & 383.6 & 83.42 & 19.29(23.1\%) \\
			& $\uparrow$ & 6.9\% & 6.5\% & 16.3\% \\
			\hline
		\end{tabular}
	}
\end{table}

\textbf{Convergence with fixed-point arithmetic.} Typically, the end-host aggregates parameters by using floating-point numbers but most current commodity switches only support fixed-point arithmetic, which may vanish parameters during the training process. Our developed FPGA board indeed can support both floating-point and fixed-point arithmetic and we are interested to explore whether NetReduce with fixed-point arithmetic would impact the training convergence. We convert the floating-point parameters to the fixed-point ones in end-hosts by keeping the original significant digits of the parameters.
\begin{figure}[t]
\centerline{\includegraphics[scale=0.6]{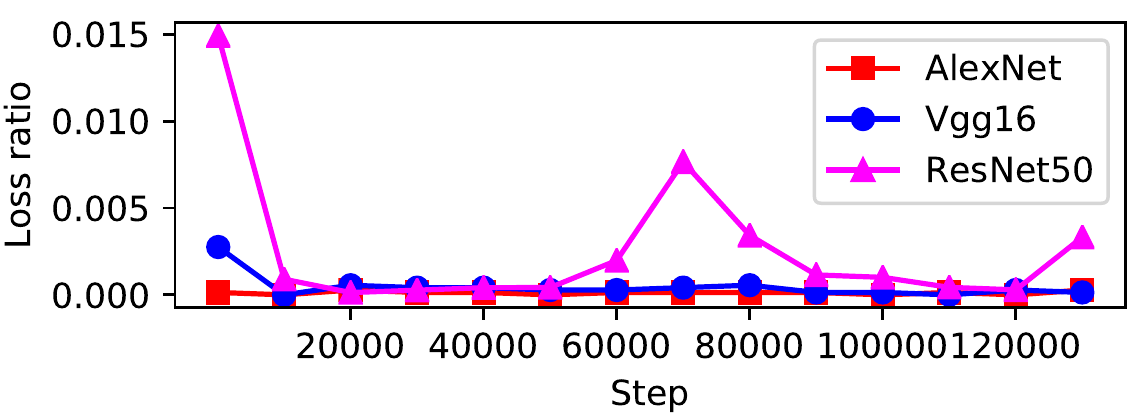}}
\caption{Absolute loss difference ratio between fixed-point NetReduce and floating-point ring all-reduce.}
\label{fig:convergence}
\end{figure}

We explore the absolute loss difference between NetReduce with fixed points and ring all-reduce with floating points, and developed a metric, $\frac{|LOSS_{inet}-LOSS_{ring}|}{LOSS_{ring}}$, as shown in Figure~\ref{fig:convergence}. Despite the initial value, the loss difference accounts for no more than 0.01\% of the baseline for AlexNet and VGG-16. ResNet-50 seems to be more sensitive to fixed-point arithmetic and the loss difference vibrates during the training process. Nevertheless, the largest loss difference accounts for less than 0.08\% of the baseline. Therefore, we can conclude that NetReduce operation with fixed-point arithmetic does not impact the training convergence.
\begin{figure}[t]
	\centerline{\includegraphics[scale=0.7]{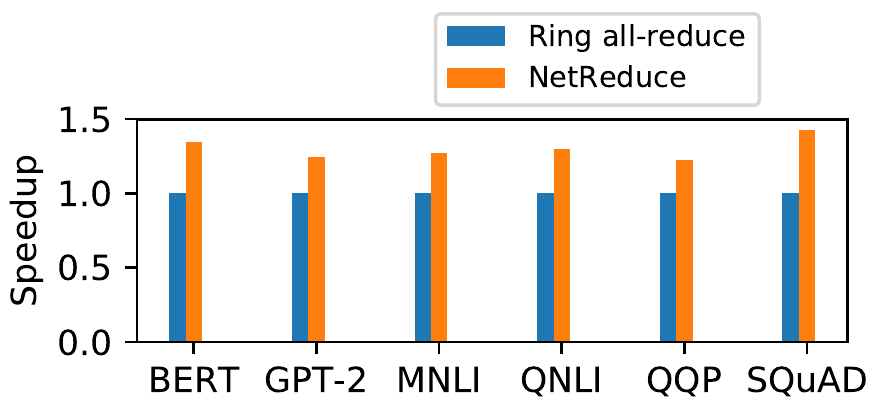}}
	\caption{Speedup of NetReduce over ring all-reduce on NLP tasks by using 6 NVIDIA GeForce RTX 2080 GPUs.}
	\label{fig:nlp}
\end{figure}

\textbf{Models other than CNN.} Besides CNN models, we also use NetReduce to pretrain and fine-tune transformer-based models for NLP jobs. We pretrain the famous BERT and GPT-2 and further fine-tunes BERT for tasks including the General Language Understanding Evaluation (GLUE) and Stanford Question Answering Dataset (SQuAD). The GLUE is a collection of tasks for evaluating natural language understanding systems. We select three corpora for the evaluation of NetReduce, MNLI, QNLI, and QQP. The SQuAD is a reading comprehension dataset for question answering, combining the 100,000 questions with over 50,000 unanswerable questions. The speedup of NetReduce over ring all-reduce on the tasks is shown in Figure~\ref{fig:nlp}, where NetReduce improves BERT pretraining, GPT-2 pretraining, GLUE-MNLI, GLUE-QUNLI, GLUE-QQP, and SQuAD by 34.6\%, 24.8\%, 27.3\%, 29.6\%, 22.2\%, 42.5\%, respectively.

%-------------------------------------------------------------------------------
\subsection{Multi-Machines Multi-GPUs Scenario}\label{subsec:eval_mmmg}
%-------------------------------------------------------------------------------

\begin{figure}[t]
	\centerline{\includegraphics[scale=0.75]{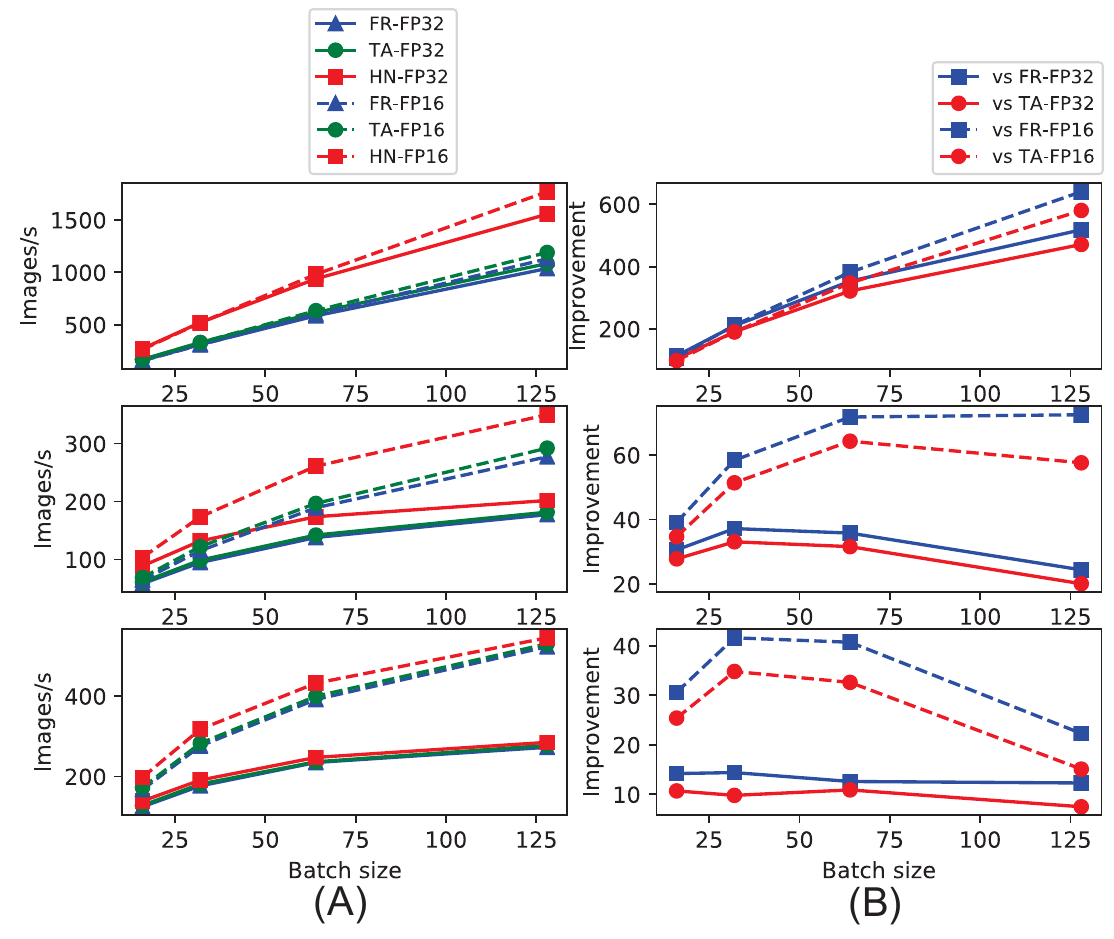}}
	\caption{Training throughput per GPU with various batch sizes and precision in multi-machines multi-GPUs scenario (4 machines each with 8 Tesla V100; FR: flat ring all-reduce; TA: Tencent all-reduce; HN: hierarchical NetReduce): (A) image processing throughput; (B) absolute throughput improvement. The $1^{st}$ to $3^{rd}$ rows refer to AlexNet, VGG-16, and ResNet-50, respectively.}
	\label{fig:mmmg}
\end{figure}
In this scenario, 4 machines equipped with 8 Tesla V100 GPUs respectively are used. We compare the training performance of hierarchical NetReduce (HN) with that of flat ring all-reduce (FR) and Tencent all-reduce (TA) as shown in Figure~\ref{fig:mmmg}. Various BSs are chosen: 32, 64, 128, and 256. The absolute improvement pattern in Figure~\ref{fig:mmmg}(B) is similar to that in Figure~\ref{fig:mmsg}(B). To sort the algorithms by training speed from fast to slow, we have the following ranking: hierarchical NetReduce, Tencent all-reduce, and flat ring. The training performance using BS=32, FP16 is summarized in Table~\ref{tab:mmmg}. Hierarchical NetReduce improves flat ring by 68.8\%, 50.7\% and 15.1\% for AlexNet, VGG-16, and ResNet-50, respectively. Compared with Tencent all-reduce, hierarchical NetReduce speeds up training by 57.9\%, 42.1\% and 12.3\% for the three models respectively. 
\begin{table}[t]
	\centering
	\caption{Training performance per GPU with BS=32, FP16 by using 4 machines each with 8 Tesla V100.}
	\label{tab:mmmg}
	\resizebox{\linewidth}{!}{
		\begin{tabular}{cc|cccc}
			\hline
			\multicolumn{2}{c|}{\multirow{2}{*}{Model}} & Flat ring & Tencent & Hierarchical \\
			& & all-reduce & all-reduce & NetReduce  \\ \hline
			AlexNet & Images/s & 307.5 & 328.8 & 519.2 \\
			(236 MB) & $\uparrow$ & 68.8\% & 57.9\% & - \\ \hline
			VGG-16 & Images/s & 115.2 & 122.2 & 173.6 \\
			(528 MB) & $\uparrow$ & 50.7\% & 42.1\% & - \\ \hline
			ResNet-50 & Images/s & 276.0 & 282.8 & 317.6 \\
			(98 MB) & $\uparrow$ & 15.1\% & 12.3\% & - \\
			\hline
		\end{tabular}
	}
\end{table}

It is reported in~\cite{jia2018highly} that Tencent hierarchical algorithm only brings performance gain for tensors with smaller sizes. For relatively larger tensors, the flat ring algorithm still outperforms the hierarchical algorithm. Recall in~\cref{subsec:mulgpu}, the communication cost consists of two items: message processing latency item with $\alpha$ and tensor transmission item with $M$. The $\alpha$ item is mostly affected by the number of GPUs participating in training, $P$. With increased $P$, the $\alpha$ item accounts for a larger proportion in flat ring all-reduce, resulting in poor scalability. Hierarchical approaches reduce the impact of $\alpha$ item by dividing a big ring into multiple small intra rings, improving the scalability. Therefore, for small tensors where the $\alpha$ item accounts for most communication costs, hierarchical approaches give superior performance. However, for big tensors where the $M$ item accounts for most communication costs and the system becomes less sensitive to $P$, hierarchical approaches bring fewer benefits.

Nevertheless, when the bandwidth of intra and inter rings fulfill certain conditions, hierarchical approaches can outperform the flat ring regardless of tensor size. Specifically, hierarchical NetReduce would always outperform flat ring if condition (~\ref{eq:condition}) holds. Considering our hardware prototype, substituting P=32 and n=8 into (~\ref{eq:condition}) gives $\frac{B_{intra}}{B_{inter}} \geq$ 2.3. Indeed intra and inter nodes being connected via NVLink and 100GbE, gives $B_{intra}$ = 150 GB/s and $B_{inter}$ = 12.5 GB/s, respectively. Therefore, in our hardware prototype, $\frac{B_{intra}}{B_{inter}}$=12 $>$ 2.3. In the next subsection, we will further explore different ratios of intra and inter rings bandwidth in the situation involving up to thousands of GPUs. 
\begin{figure}[t]
	\centerline{\includegraphics[scale=0.97]{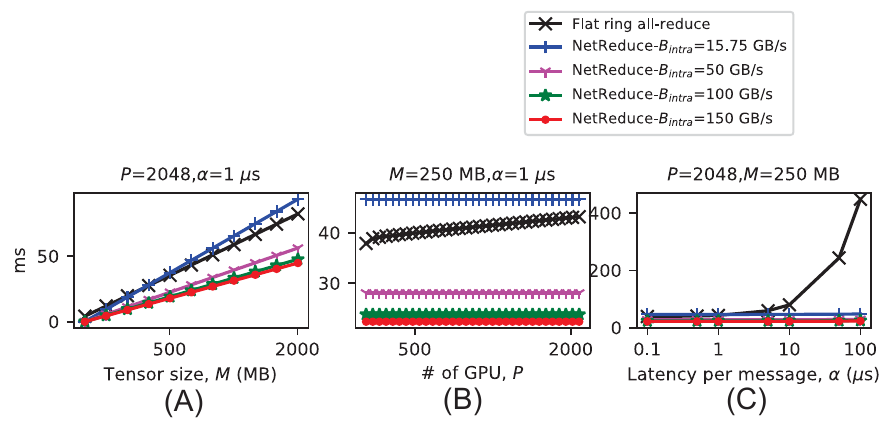}}
	\caption{Communication cost taken by a single machine for parameter synchronization ($n$=8, $B_{inter}$=12.5 GB/s): (A) v.s. Tensor size ($M$); (B) v.s. number of GPU ($P$); (C) v.s. latency per tensor ($\alpha$).}
	\label{fig:simulation}
\end{figure} 

%-------------------------------------------------------------------------------
\subsection{Large-Scale System Simulation}\label{subsec:simulation}
%-------------------------------------------------------------------------------

Recall Eqs.(~\ref{eq:fr}) to (~\ref{eq:deltafr}) back in~\cref{subsec:mulgpu}, the major factors affecting communication cost includes processing latency per message ($\alpha$), tensor size ($M$), number of GPUs ($P$), intra ring bandwidth inside one single machine ($B_{intra}$). We conduct simulations on these metrics to explore what effects they would bring to the system.

The simulation results are shown in Figure~\ref{fig:simulation}, where $n=8$ and $B_{inter}=$12.5 GB/s by using 100 GbE NIC. $B_{intra}$ are chosen from 15.75 GB/s (PCIe) to 150 GB/s (NVLink). Figure~\ref{fig:simulation}(A), (B), and (C) show the result of communication time v.s. $M$, $P$, and $\alpha$, respectively. As expected, when increasing $B_{intra}$, hierarchical NetReduce consumes less time as shown in Figure~\ref{fig:simulation}, leading to larger throughput gain. However, the gain does not always rise that fast. It saturates when hitting a certain threshold of the bandwidth, e.g., increasing $B_{intra}$ from 100 GB/s to 150 GB/s does not bring that much benefit as increasing from 50 GB/s to 100 GB/s, as shown in Figure~\ref{fig:simulation}(A) and (B).       

In Figure~\ref{fig:simulation}(A), $P$ and $\alpha$ are fixed to 2048 and 1 $\mu s$ respectively. When $B_{intra}$=15.75 GB/s, hierarchical NetReduce is only better than flat ring all-reduce for tensors smaller than a threshold around 130 MB. This is consistent with the observation reported in~\cite{jia2018highly} that the benefit brought by hierarchy cannot cover the tensor transmission cost of a relatively larger $M$. Nevertheless, this can be indeed overcome by increasing the value of $B_{intra}$, e.g., to 50 GB/s, 100 GB/s, and 150 GB/s as shown in Figure~\ref{fig:simulation}(A) (See the explanation in~\cref{subsec:eval_mmmg}).   

The communication cost of hierarchical NetReduce is independent of $P$ as shown in Figure~\ref{fig:simulation}(B), where $M$ and $\alpha$ are fixed to 250 MB and 1 $\mu s$ respectively. Flat ring all-reduce takes more latency when more GPUs participating in the training job as increased $P$ leads to a higher number of transmissions. On the contrary, communication time in NetReduce is a constant regardless of the value of $P$. Similarly, as shown in Figure~\ref{fig:simulation}(C), flat ring all-reduce is more easily affected by increased $\alpha$. This is because the coefficient $2(P-1)$ amplifies the impact of $\alpha$ (Eq.(\ref{eq:fr})) while NetReduce reduces $2(P-1)$ to 1 (Eq.(\ref{eq:nh})).

In summary, hierarchical NetReduce can cover the transmission cost of large $M$ by using larger intra bandwidth. With increased $P$ and $\alpha$, hierarchical NetReduce shows better scalability than flat ring all-reduce as the performance of NetReduce is independent of the number of GPUs.

%-------------------------------------------------------------------------------
\section{Conclusion}\label{sec:conclusion}
%-------------------------------------------------------------------------------

In this paper, we present NetReduce, a novel RDMA-compatible in-network reduction architecture to accelerate distributed DNN training. Compared with existing designs, NetReduce maintains reliable connections between end-hosts in the Ethernet to avoid implementing high-cost network protocol stack in the switch. The prototype implemented by using FPGA is an out-of-box solution without modifying commodity devices such as NICs or switches. NetReduce improves the training up to 1.7x and 1.5x for CNN-based CV and transformer-based NLP tasks, respectively. We find that NetReduce with fixed-point arithmetic does not impact the training convergence. Simulations on large-scale systems indicate the superior scalability of NetReduce to the state-of-the-art ring all-reduce.

%-------------------------------------------------------------------------------
\bibliographystyle{plain}
\bibliography{Ref}

%%%%%%%%%%%%%%%%%%%%%%%%%%%%%%%%%%%%%%%%%%%%%%%%%%%%%%%%%%%%%%%%%%%%%%%%%%%%%%%%
\end{document}